# Origin of indirect optical transitions in few-layer MoS$_2$, WS$_2$ and WSe$_2$

*Weijie Zhao[a,b], R. M. Ribeiro[b,c], Minglin Toh[a,b], Alexandra Carvalho[a,b], Christian Kloc[d], A. H. Castro Neto[a,b], Goki Eda[a,b,e,*]*

[a] Department of Physics, National University of Singapore, 2 Science Drive 3, Singapore 117542

[b] Graphene Research Centre, National University of Singapore, 6 Science Drive 2, Singapore 117546

[c] Center of Physics and Department of Physics, University of Minho, PT-4710-057, Braga, Portugal

[d] School of Materials Science and Engineering, Nanyang Technological University, N4.1 Nanyang Avenue, Singapore 639798

[e] Department of Chemistry, National University of Singapore, 3 Science Drive 3, Singapore 117543

[*] E-mail: g.eda@nus.edu.sg

*Abstract*

It has been well established that single layer MX$_2$ (M=Mo,W and X=S,Se) are direct gap semiconductors with band edges coinciding at the K point in contrast to their indirect gap multilayer counterparts. In few-layer MX$_2$, there are two valleys along the Γ-K line with similar energy. There is little understanding on which of the two valleys forms the conduction band minimum (CBM) in this thickness regime. We investigate the conduction band valley structure in few-layer MX$_2$ by examining the temperature-dependent shift of indirect exciton PL. Hihgly anisotropic thermal expansion of the lattice and corresponding evolution of the band structure result in distinct peak shift for indirect transitions involving the K and Λ (midpoint along Γ-K) valleys. We identify the origin of the indirect emission and concurrently determine the relative energy of these valleys. Our results show that the two valleys compete in energy in few-layer WSe$_2$.

Electronic structure of two-dimensional (2D) crystals such as graphene and $MoS_2$ depends strongly on thickness and stacking sequence of individual layers[1-7]. Single layers are most distinct in that their thickness is in its ultimate limit. Since the interaction between layers is much weaker than the in-plane interactions, the low energy electronic properties of multilayers are highly sensitive to the number layers and their relative orientation. For instance, monolayer graphene exhibits linear dispersion at low energies while Bernal-stacked bilayer graphene shows parabolic dispersion.[1, 2, 8] Similarly, direct band gap of single layer $MoS_2$ is in contrast to indirect band gap of their multilayer counterparts[4, 7, 9, 10]. In bilayers and thicker multilayers, interlayer interaction, geometrical confinement, and crystal symmetry play a collective role in defining their electronic structures.[1-7] Correspondingly, stark differences are observed in electrical and optical properties of these materials in the single- to few-layer thickness regime. [5, 6, 11]

The band structure of $MoS_2$ and its isoelectronic compounds of the group 6 transition metal dichalcogenide (TMD) family, such as $MoSe_2$, $WS_2$, and $WSe_2$, is distinctly different from that of graphene. The conduction band valleys are located at the six corners of the Brillouin zone (K/K' points) and at midpoints along high symmetry lines Γ-K and M-K. The valence band hills are located at the Γ point as well as at the K/K' points where the band is split due to spin-orbit coupling. Recent reports on valley polarization in $MX_2$ single layers[3, 12-15] highlight the unique band structure of these materials and its exploitation.

The indirect-to-direct crossover in these $MX_2$ compounds (M=Mo,W and X=S,Se) results from local shift of valence band hills and conduction band valleys in the Brillouin zone[4, 7, 9, 10, 16-19]. In single layers, the conduction band minimum (CBM) and valence band minimum (VBM) coinciding at the K point, making them direct gap semiconductors. In multilayers, the valence band hill at the Γ point is raised above the hill at the K point. This behavior was predicted by calculations[7, 16-18, 20-22] and confirmed in early photoemission experiments on $WS_2$ monolayer grown by van der Waals epitaxy[23]. Similarly, the conduction band valley at the Λ point (midpoint between K and Γ) shifts downward in the presence of interlayer interaction as indicated by calculations[7, 16, 17, 20-22]. In contrast, the states near the K point are comparatively less susceptible to the number of layers[4, 7, 9, 10].

While it is well established that few-layer $MX_2$ are indirect gap semiconductors, there has been little consensus on the location of the CBM, which is fundamentally important in understanding charge transport and optical effects. In few-layer $MX_2$, the conduction band valleys of the K and Λ points compete in energy. For bilayer $MoS_2$ and $WS_2$, some calculations[16, 17, 21, 22, 24] show that the CBM is located at the K point while others[7, 16, 17, 20] show that it is at the Λ point. The discrepancy is partly due to the fact that some calculations neglect spin-orbit interactions. In bilayer $WSe_2$, the two valleys are expected to be nearly degenerate[25]. Degeneracy of these valleys indicates increased probability for intervalley scattering of the carriers. Despite the fundamental significance, experimental evidence for the energetics of these valleys is currently lacking.

Here, we investigate the competition of conduction band valleys in few-layer $MX_2$ by studying radiative indirect optical transitions and their temperature-dependence. Indirect exciton recombination in few-layer $MX_2$ can preferentially occur via either K or Λ valleys depending on the optical gap. Considering the binding energy of the two indirect excitons and origin of the indirect transitions, it is possible to identify which valley represents the CBM. In our experiment, temperature is used as a knob to continuously tune the interlayer interaction and corresponding evolution of the band structure i.e. upward or downward shifts of the Λ valley and Γ hill. We compare the temperature-dependent shift in the optical transition energies to that predicted by DFT calculations and conclude the origin of the indirect transition. Our results indicate that the CBM is located at the Λ point for bilayer $MoS_2$ and $WS_2$ while it is at the K point for bilayer $WSe_2$.

**Results and Discussion**

We conducted fully relativistic DFT calculations of the electronic band structure of bilayer $MoS_2$, $WS_2$, and $WSe_2$ as shown in Figure 1 (See Methods for details). All materials exhibit VBM at the Γ point and a second highest valence band hill at the K point. These results show that CBM is located at the K point for $MoS_2$ and $WSe_2$ while it is at the Λ point for $WS_2$. The energy difference between the Λ and K valley minima is 100, 85 and 55 meV for $MoS_2$, $WS_2$, and $WSe_2$, respectively. These results suggest that excited electrons will relax to the valence band via K→Γ indirect transition for $MoS_2$ and $WSe_2$ whereas Λ→Γ transition is favored for $WS_2$ (dashed arrows in Figure 1). In real systems, however, excitonic and polaronic effects are expected to play a major role, leading to deviations in the relaxation pathways predicted by the static ground state picture[11, 18, 21, 22, 26]. Thus, DFT results may not be directly correlated with experimental observations.

Indirect band gap photoluminescence (PL) was previously observed in bilayer $MoS_2$, $WS_2$, and $WSe_2$. However, precise origin of the peak remains elusive since the two indirect optical transitions K→Γ and Λ→Γ are indistinguishable from the emission spectrum. In order to determine the origin of indirect transition, we consider temperature as the knob to continuously tune the band structure and study its trend. Thermal expansion causes the interlayer spacing to increase with temperature[27, 28], thus reducing the interaction between the layers. At sufficiently high temperatures, each layer in the stack is only weakly coupled such that it behaves like an isolated monolayer. Thus, the band structure is expected to evolve towards indirect-to-direct crossover point with temperature[10, 29]. It should be noted that the in-plane thermal expansion, which is significantly smaller compared to c-axis expansion, also has distinct effects on the band structure.

In Figure 2, we show our DFT results on the variation of electronic transition energies $E_{K \to K}^{elec}$, $E_{K \to \Gamma}^{elec}$, and $E_{\Lambda \to \Gamma}^{elec}$ (The superscript denotes that this is an electronic gap) as a function of lattice expansion. Increase in the in-plane lattice constant by Δa leads to reduction of direct and indirect transition energies at a different rate (Figure 2a and d) in agreement with the previous studies[16, 18, 19, 24, 30, 31]. The changes are more pronounced for transitions involving the K point. Changes in interlayer spacing have an opposite effect on the transition energies (Figure 2b and e). Increase in interlayer spacing by Δc leads to increase in $E_{\Lambda \to \Gamma}^{elec}$ and $E_{K \to \Gamma}^{elec}$ whereas $E_{K \to K}^{elec}$ transition energy remains nearly unchanged. This trend leads to indirect-to-direct gap crossover at a large interlayer spacing.[29]

Similar to reducing the number of layers, increasing interlayer spacing results in upshift of the Λ valley and downshift of the Γ hill. These regions of the band structure describe wave functions that decay slowly outside the plane of $MX_2$ sandwich. In contrast, K valleys and hills define the wave functions that are localized around the transition metal atoms[7, 11]. Thus, the K valleys/hills are significantly less susceptible to interlayer spacing. These general trends are the same for $MoS_2$, $WS_2$, and $WSe_2$ (See Supporting Information for the results on $WS_2$).

Next, we examine the effects of thermal expansion. Figures 2c and f shows the changes in the transition energies as a function of temperature in the range of 123 K and 473 K. Experimentally measured thermal coefficients for bulk $MoS_2$ and $WSe_2$[27, 28] were used for our calculations. According to recent reports, thermal expansion coefficients of few-layer $MX_2$ are expected to be comparable to those of their bulk counterparts[10]. Our results show that $E_{K \to K}^{elec}$ and $E_{K \to \Gamma}^{elec}$ decrease with temperature at a similar rate, indicating that the trend is mainly due to in-plane lattice expansion. In contrast, $E_{\Lambda \to \Gamma}^{elec}$ increases with temperature, indicating that the effect due to out-of-plane thermal expansion outplays that due to in-plane thermal expansion. The trends are more pronounced for $WSe_2$, which exhibits larger thermal expansion coefficients compared to MoS$_2$.[27, 28] The opposite temperature dependence of $E_{K \to \Gamma}^{elec}$ and $E_{\Lambda \to \Gamma}^{elec}$ found here provides a clue to the origin of the experimentally observed indirect optical transition in these materials.

We now turn to our experimental observation of the PL spectra and their temperature dependence. Temperature-dependent PL spectra were obtained for mechanically exfoliated few-layer MX$_2$ samples on quartz and SiO$_2$/Si substrates. Figure 3a-c shows the PL spectra of bilayer MoS$_2$, WS$_2$ and WSe$_2$ obtained at different temperatures from 123K to 473K. Direct and indirect exciton peaks are labeled as A/B and I, respectively. Each emission peak was fitted with Gaussian and their temperature-dependent shift is plotted in Figures 3d-f. The direct exciton peaks A and B redshift with increasing temperature in agreement with the previous reports[10, 32] and the DFT results discussed above. The temperature coefficient of these peaks for all materials was found to be about -0.3 meV/K. On the other hand, contrasting differences were observed in the temperature dependence of the indirect emission peaks. Interestingly, the temperature coefficient of the indirect exciton peak is positive for MoS$_2$ and WS$_2$ but negative for WSe$_2$ for most temperature ranges.

According to the DFT results shown in Figure 2c and f, the temperature coefficient is positive for $E_{\Lambda \to \Gamma}$ and negative for $E_{K \to \Gamma}$. Thus, our experimental results indicate that the indirect emission originates from $\Lambda \to \Gamma$ transition in MoS$_2$ and WS$_2$ but $K \to \Gamma$ transition in WSe$_2$. These experimental observations are consistent with the DFT results for WS$_2$ and WSe$_2$ where the CBM is located at the $\Lambda$ and K point, respectively. However, the experimental results for bilayer MoS$_2$ are unexpected from DFT results, which predict that the CBM is located at the K point. The discrepancy may be due to different binding energies of the indirect gap excitons or polaronic effects. We studied Stokes shift due to thermalization of carriers as a potential origin of the observed temperature dependence (See Supporting Information for details). However, contribution from Stokes shift was found to be minor compared to the overall shift.

We note that exciton-phonon interaction also results in shift of emission peak energies. General trend is that the energy and lifetime of excitons decay linearly with temperature at above the Debye temperature $T_D$ and remain constant below $T_D$ (See Supporting Information for details). The phonon effects appear to enhance the temperature dependent shift beyond the prediction by the ground state DFT results in agreement with the results by Tongay et al.[10]

The indirect emission peak shift in $WSe_2$ shows an unexpected trend below 173 K where the temperature dependence is reversed (Fig 3f). We attribute this to emergence of a new indirect emission peak which originates from Λ→Γ transition. Since the energy difference between the Λ and K valley minimum is as small as 55 meV for bilayer $WSe_2$ according to the DFT results, we expect to observe crossover of Λ→Γ and K→Γ indirect emission peaks, which exhibit opposite temperature dependence.

In order to verify the origin of indirect emission in $WSe_2$, we studied the temperature-dependent PL spectrum of thicker multilayer $WSe_2$ (Figure 4a-c). The direct emission peak for the 2, 3, 4, and 8L $WSe_2$ shows similar temperature dependence with a temperature coefficient of 0.35 meV/K. The indirect emission peak region of the spectra (800~1000 nm) indicates presence of two distinct peaks that exhibit opposite temperature dependence. At temperatures below 173 K, a peak (denoted as $I_1$ in Figure 4) is found to blueshift with temperature whereas at above 373 K, a peak with an opposite trend (denoted as $I_2$ in Figure 4) is observed. It can be seen that these are independent peaks because, at intermediate temperatures, they merge to form a broad peak that can be deconvoluted with two Gaussians. The two peaks are clearly observable for 8L $WSe_2$ at 273 K (0 ℃) (Figure 4c). The temperature dependence of $I_1$ and $I_2$ peaks is shown in Figure 4d-f.

Observation of two emission peaks $I_1$ and $I_2$ with opposite temperature dependence in the indirect emission peak region strongly suggests that they originate from K→Γ and Λ→Γ indirect transitions, respectively. These results indicate that K and Λ valleys compete in energy and their relative shift with temperature leads to changes in the favored indirect transition.

The calculated band structure for 3L, 4L and bulk $WSe_2$ (Figure 5) shows that the CBM is located at the K point in contrast to the case of 2L $WSe_2$. In Figure 5b, we compare experimentally observed optical transition energies ($E^{opt}$) in $WSe_2$ as a function of the number of layers. The $E_{\Lambda\to\Gamma}^{opt}$ values are room temperature extrapolation of the temperature dependence. It can be seen that $E_{\Lambda\to\Gamma}^{opt}$ decreases more rapidly with increasing number of layers compared to $E_{K\to\Gamma}^{opt}$ and the crossover of the transition occurs between 2 and 3 layers in agreement with the DFT results. The energy difference between $E_{\Lambda\to\Gamma}$ and $E_{K\to\Gamma}$ also shows good agreement between experimental and calculated results (Figure 5c).

The optical gap that we obtain experimentally is offset from the electronic gap by the exciton binding energy. From the DFT results, the electron effective mass at the Λ valley is found to be larger than that at the K valley by 20~30 % (See Supporting Information for details). Correspondingly, excitons involving the Λ valley are more strongly bound. Thus, when the lowest energy optical indirect transition is K→Γ (as it is the case of 2L $WSe_2$) then it confirms that the CBM is at the K point. When the lowest energy optical indirect transition is Λ→Γ, then accurate determination of the location of CBM requires the binding energy of the excitons to be precisely known. For 8L $WSe_2$, $E_{K\to\Gamma}^{opt}$ - $E_{\Lambda\to\Gamma}^{opt}$ ~ 100 meV is sufficiently larger than the estimated difference in the indirect exciton binding energies (See Supporting Information for details) and thus it may be concluded that the CBM occurs at the Λ point.

In summary, we explore the conduction band valley structure of few-layer $MX_2$ by close examination of temperature-dependent indirect exciton emission peaks. Anisotropic thermal expansion of the lattice and corresponding evolution of the band structure provides a unique fingerprint for different indirect transitions. We identify the favored valley in the indirect transition depends on the material, the number of layer, and temperature. Our findings shed light on the valley structure and its potential implications on the electrical and optical properties of these materials.

Methods

Band structure calculation: The DFT calculations were performed using the open source code Quantum ESPRESSO[33]. We used norm conserving, full relativistic pseudopotentials with nonlinear core-correction and spin-orbit information to describe the ion cores. The pseudopotentials from the Quantum ESPRESSO distribution were used when available, subject to tests for typical systems. Otherwise, the pseudopotentials were produced using the atomic code by A. Dal Corso, in the Quantum ESPRESSO distribution. The exchange correlation energy was described by the generalized gradient approximation (GGA), in the scheme proposed by Perdew-Burke-Ernzerhof (PBE)[34], and with the semi-empirical dispersion term of Grimme[35, 36]. The cutoff energy used was 25 Ry, and it was confirmed that the bandstructures were converged by comparison with 50 Ry calculations. The Brillouin-zone (BZ) was sampled for integrations according to the scheme proposed by Monkhorst-Pack[37], with grids of 8x8x1 or 12x12x1 k-points. Samples were modeled in a slab geometry by including a vacuum region of about 90 Bohr in the direction perpendicular to the surface. The in-plane lattice parameter was determined by minimization of the total energy, for the monolayers. For the bulk crystals, we used the experimental value of the ratio of lattice parameters $c/a$[38]. This inter-layer distance was also used as a starting point for multi-layer calculations. Subsequently, the atoms were relaxed until all forces were less than $10^{-3}$ atomic units.

Experiment: Crystals of 2H-$WS_2$ and 2H-$WSe_2$ were grown by chemical vapor transport (CVT) using iodine as the transport agent. Natural 2H-$MoS_2$ crystals were obtained from SPI Supplies. The crystals were mechanically exfoliated and deposited on quartz and $SiO_2$/Si substrates(300nm $SiO_2$). The number of layers for the exfoliated samples was verified by atomic force microscope (AFM), Raman spectroscopy and optical contrast method (i.e. differential reflectance).[9, 39] Differential reflectance measurements were conducted with a tungsten-halogen lamp coupled with a Raman spectrometer. The beam spot from the tungsten-halogen lamp was focused on the exfoliated monolayer and multilayer samples with a diameter of about 3 micron. PL spectra were obtained by a back scattering geometry with a 473 nm excitation laser at intensities of around 30 μW. A Linkam cryostat with an optical window was used for a temperature control range of around 123 K to 573 K for the temperature-dependent PL measurements. The samples were protected under nitrogen ambient for all the temperature range in order to avoid any oxidation of them. The temparature-dependent PL for each sample was performed for several circles in order to confirm the reproducibility. The obtained PL spectra were carefully corrected with the correction factors obtained by using the standard tungsten-halogen lamp (detailed in the Supporting Information).

Acknowledgement

G.E acknowledges Singapore National Research Foundation for funding the research under NRF Research Fellowship (NRF-NRFF2011-02) and Graphene Research Centre. R.M.R is thankful for the financial support by FEDER through the COMPETE Program and by the Portuguese Foundation for Science and Technology (FCT) in the framework of the Strategic Project PEST-C/FIS/UI607/2011 and grant nr. SFRH/BSAB/1249/2012.

Figure 1

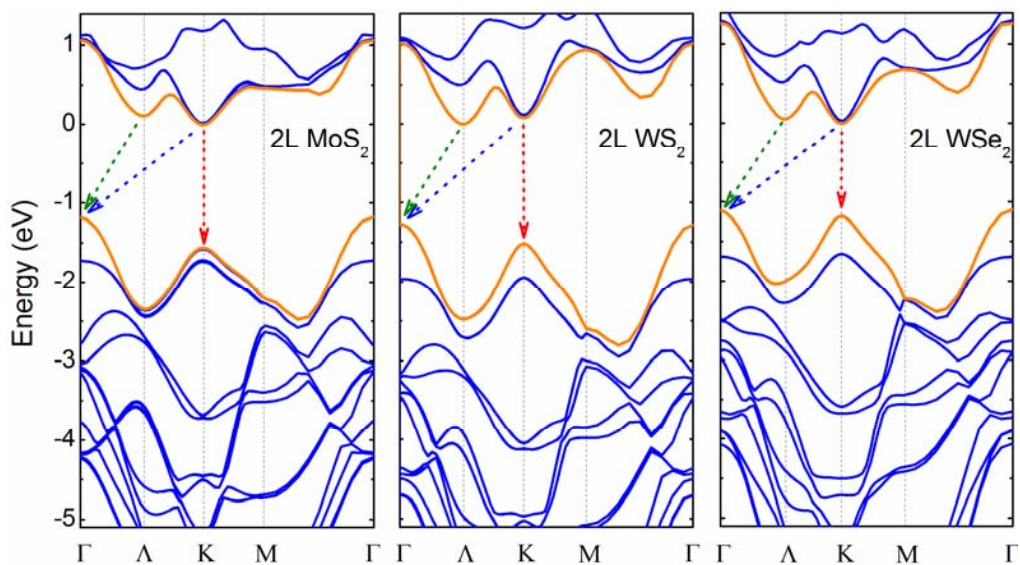

Calculated energy band structure of the bilayer MoS$_2$, WS$_2$ and WSe$_2$. The dashed arrows indicate the possible radiative recombination pathways (K→K, K→Γ, and Λ→Γ) for photoexcited electron-hole pairs. The bands forming the conduction band minimum and valence band maximum are indicated in orange.

Figure 2

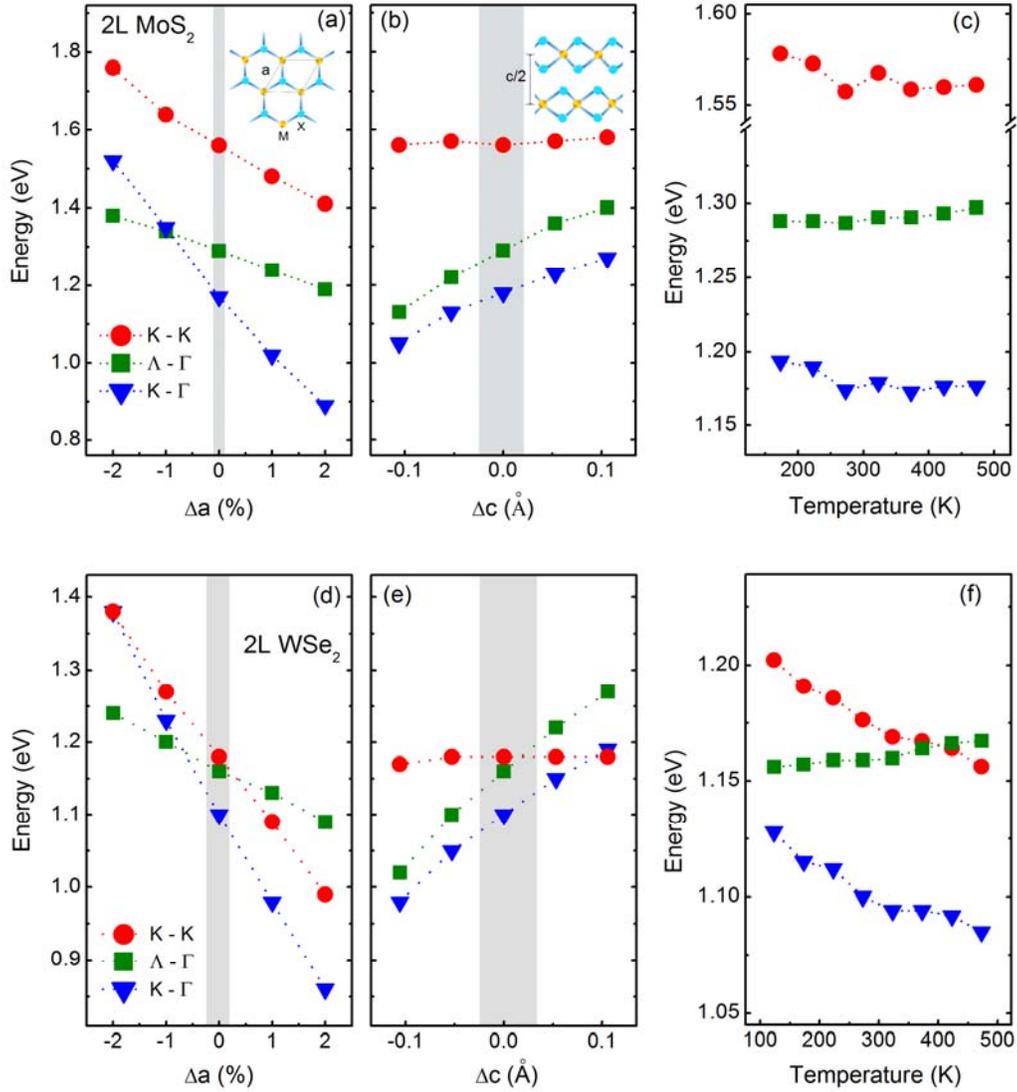

The evolution of direct and indirect band gaps of (a-c) 2L $MoS_2$ and (d-f) $WSe_2$ as a function of change in in-plane lattice constant Δa (a and d), change in out-plane lattice constant Δc (b and e), and temperature. The gray region in (a), (b), (d) and (e) indicate the changes that occur in the temperature range of -123K to 473K. The dotted lines are guides for the eye. Inset of (a) and (b) are schematic of the lattice showing the in-plane lattice constant 'a' and out-plane lattice constant 'c', respectively. M and X represent the transition metal atoms and chalcogen atoms, respectively.

Figure 3

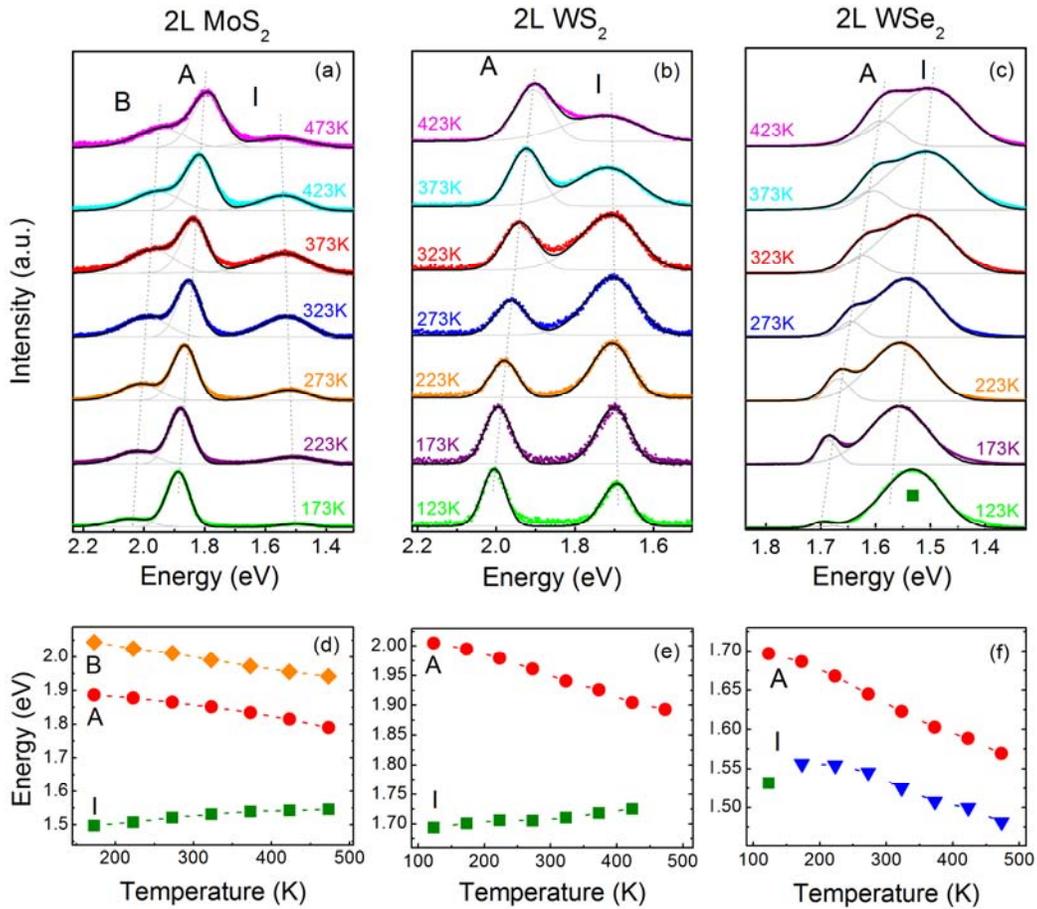

PL spectra of 2L $MoS_2$ (a), $WS_2$ (b) and $WSe_2$ (c) in the temperature range of 123K to 473K. The colored plots are the experimental data. The PL Spectra were fitted with Gaussian peaks (grey lines). The black lines are the fitting results. The corresponding evolutions of the direct and indirect emissions are extracted in (d), (e) and (f), respectively. A and B indicate the direct emission from direct band gaps, while I indicates the indirect emission from indirect band gap. The dotted/dashed lines are guides for the eye.

Figure 4

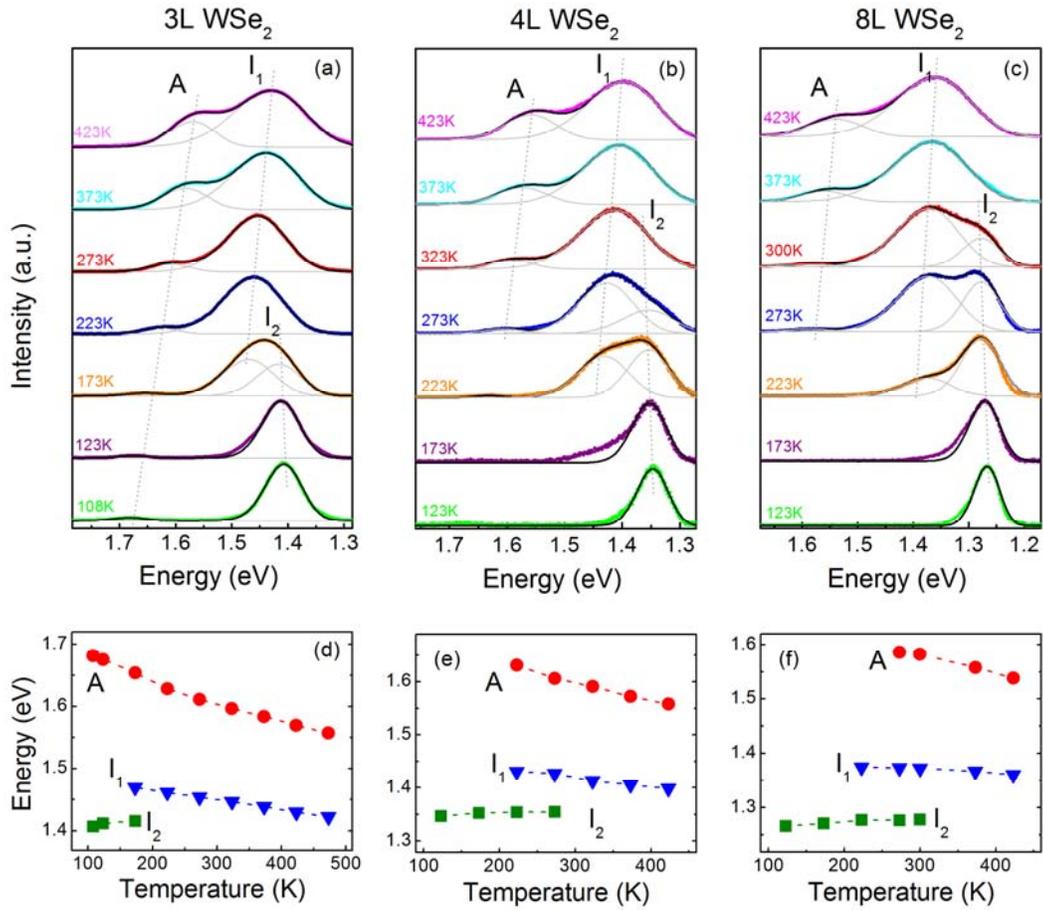

The temperature dependent PL spectra of a 3L (a), 4L (b) and 8L (c) WSe$_2$. The colored lines are the experimental data. The PL spectra were fitted with Gaussian peaks (grey lines). The black lines are the fitting results. The corresponding evolutions of the direct and indirect emissions are extracted in (d), (e) and (f), respectively. I$_1$ and I$_2$ indicate the indirect emission, while A indicates the direct emission. The dotted/dashed lines are guides for the eye.

Figure 5

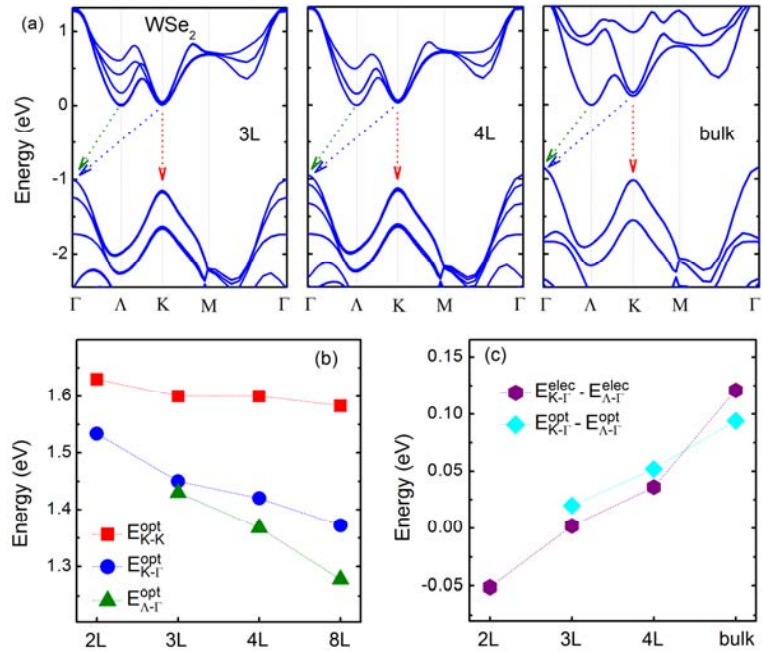

(a) Calculated electronic band structure of the 3L, 4L and bulk WSe$_2$. The dashed arrows indicate the possible radiative recombination pathways (K→K, K→Γ, and Λ→Γ) for photoexcited electron-hole pairs. (b) The peak position of the direct and indirect emission for the 2-4L and 8L WSe$_2$ at room temperature (300 K). The data for the 3L, 4L and 8L WSe$_2$ are extrapolated from Figure 4d-f. (c) The comparison of the energy difference between K→Γ and Λ→Γ transitions for the calculated electronic and experimental optical band gaps. The dotted lines in (c) are guides for the eye.